# The Role of Deep Learning in Financial Asset Management: A Systematic Review

Pedro Dias Reis [a, b]

0009-0000-2912-8364

corresponding author

up200602922@edu.fep.up.pt

Ana Paula de Sousa Freitas Madureira Serra [a]

0000-0003-4312-0451

aserra@fep.up.pt

João Manuel Portela da Gama [a, b]

0000-0003-3357-1195

joao.gama@inesctec.pt

[a]University of Porto, School of Economics and Management and Centre for Economics and Finance (cef.up), R. Dr. Roberto Frias 46, 4200-464 Porto, Portugal

[b]INESCTEC, Campus da Faculdade de Engenharia da Universidade do Porto, R. Dr. Roberto Frias, 4200-465 Porto, Portugal

This work was carried out within the scope of the research project funded by *Fundação para a Ciência e a Tecnologia,* grant number 2023.01070.BD.

The authors declare that there are no conflicts of interest regarding the publication of this review. The research was conducted independently, and no external funding or financial relationships influenced the study design, data collection, analysis, or conclusions.

# Abstract

This review systematically examines deep learning applications in financial asset management. Unlike prior reviews, this study focuses on identifying emerging trends, such as the integration of explainable artificial intelligence (XAI) and deep reinforcement learning (DRL), and their transformative potential. It highlights new developments, including hybrid models (e.g., transformer-based architectures) and the growing use of alternative data sources such as ESG indicators and sentiment analysis. These advancements challenge traditional financial paradigms and set the stage for a deeper understanding of the evolving landscape. We use the Scopus database to select the most relevant articles published from 2018 to 2023. The inclusion criteria encompassed articles that explicitly apply deep learning models within financial asset management. We excluded studies focused on physical assets. This review also outlines our methodology for evaluating the relevance and impact of the included studies, including data sources and analytical methods. Our search identified 934 articles, with 612 meeting the inclusion criteria based on their focus and methodology. The synthesis of results from these articles provides insights into the effectiveness of deep learning models in improving portfolio performance and price forecasting accuracy. The review highlights the broad applicability and potential enhancements deep learning offers to financial asset management. Despite some limitations due to the scope of model application and variation in methodological rigour, the overall evidence supports deep learning as a valuable tool in this field. Our systematic review underscores the progressive integration of deep learning in financial asset management, suggesting a trajectory towards more sophisticated and impactful applications. Future research should explore the capabilities of these models for diverse financial contexts.

# 1 Introduction

Finance is a highly varied area of study within economics, covering a huge array of topics, such as asset and portfolio management, risk assessment, identification of fraudulent activities, and financial oversight. In recent years, integrating advanced learning methodologies, particularly deep learning (DL), has become increasingly significant in areas such as portfolio management and risk assessment. This paper reviews the application of DL models in financial asset management, focussing on their contributions to various financial decisions within financial markets.

The intersection of DL and finance has seen notable advancements, especially in improving the precision of financial forecasts, identifying complex patterns in market data, and improving risk management strategies. We have identified 15 articles (Table 1) that explore this intersection. Surveys on general applications of DL to financial markets are relatively common, with five review articles published over the last five years. Price forecasting and time series prediction topics account for six of the specific review articles. There is only one review of asset pricing. Lastly, portfolio management is highlighted in three review papers. In most cases, the publishing journals were primarily computer science outlets. However, these reviews primarily focus on standard DL techniques like LSTMs and CNNs, with limited exploration of newer advancements, such as transformer models or their applications in multi-asset portfolios. Additionally, they seldom address explainable AI methodologies, which are critical for regulatory compliance and practitioner adoption. By incorporating these recent trends, this study aims to offer a more forward-looking perspective on DL applications.

*Table 1 - List of review articles between 2019 and 2024, including general surveys and specific financial areas. General Survey (GS), Asset Management (AM), Price Forecasting (PF), Asset Pricing (AP), Portfolio Management (PM), Stock Market (SM), Exchange Rates Forecasting (FX), Bonds (BD), Interest Rate Forecasting (IR), Cryptocurrencies (CC), Commodity Prices (CP), Derivatives (DV), and Real Estate Investment Trusts (RE).*

| Author | Period studied | No. of articles | Area | Market |
| --- | --- | --- | --- | --- |
| Henrique et al. (2019) | 1991-2017 | 547 | PF | SM |
| Nti et al. (2020) | 2007-2018 | 122 | PF | SM |
| Durairaj & Mohan (2019) | 1999–2019 | 34 | PF | SM, FX, IR, DV |
| Weigand (2019) | 1994-2018 | 49 | AP | SM, IR, DV, RE |
| Emerson et al. (2019) | 2015-2018 | 55 | PM | SM |
| Ozbayoglu et al. (2020) | 2015-2020 | 144 | GS | SM, CC, DV |
| J. Huang et al. (2020) | 2014-2018 | 40 | GS | SM, FX, CP |
| Sezer et al. (2020) | 2005–2019 | 140 | PF | SM, FX, CP |
| Bustos & Pomares-Quimbaya (2020) | 2014-2018 | 53 | PF | SM |
| Mirete-Ferrer et al. (2022) | 2015–2021 | 60 | AM | SM |
| Jiang (2021) | 2017-2019 | 124 | PF | SM |
| Singh et al. (2022) | 2014-2021 | 64 | GS | SM, FX, DV, BD |
| Olorunnimbe & Viktor (2023) | 2018-2020 | 35 | AM | SM |
| Nazareth & Reddy (2023) | 2015-2023 | 126 | PM | SM, CC, FX |
| Dakalbab et al. (2024) | 2015-2023 | 143 | GS | SM, FX, CC, CP |

Henrique et al. (2019) and Nti et al. (2020) provide comprehensive surveys on price forecasting in the stock market, covering articles from 1991 to 2017 and 2007 to 2018, respectively. They both conclude that DL models significantly outperform traditional forecasting methods by capturing complex patterns in financial data. Durairaj & Mohan (2019) expanded the scope to include foreign exchange, interest rates, and derivatives, reinforcing the superior accuracy of DL models but also noting high computational demand and the risk of overfitting.

Weigand (2019) focus on empirical asset pricing, emphasising the potential of machine learning to measure stock market risk premiums. However, Weigand cautioned against overreliance on these models due to possible overfitting and data mining biases. Emerson et al. (2019) and Olorunnimbe & Viktor (2023) explore the literature on portfolio management. They highlighted the potential of reinforcement learning to adapt to market changes dynamically, thus optimising returns and managing risks more effectively than traditional methods.

Ozbayoglu et al. (2020) and J. Huang et al. (2020) provide general surveys spanning various financial markets, including cryptocurrencies, commodities, and bonds. They identify the critical need for improved model interpretability and robust validation techniques to ensure the reliability of DL and ML models. They also point out some of the ethical and regulatory challenges posed by the widespread use of these technologies.

Sezer et al. (2020) and Bustos & Pomares-Quimbaya (2020) reiterate the effectiveness of DL models in price forecasting, while Mirete-Ferrer et al. (2022) focus on financial asset management, confirming the superiority of these models in financial forecasting. They highlight the necessity for skilled model tuning and address the challenges of high computational demands and data snooping.

Jiang (2021) and Nazareth & Reddy (2023) underscore the importance of adding alternative data sources, such as sentiment analysis and macroeconomic indicators, into DL models to enhance predictive performance. These studies emphasise that alternative data could provide additional insights and improve forecast accuracy. However, they also note the ongoing challenges of model complexity and interpretability.

Singh et al. (2022) and Dakalbab et al. (2024) further stress the need for integrating traditional models with advanced DL and ML techniques. They advocate for developing robust validation methods to ensure models' generalisation and reliability. Additionally, they discuss the ethical considerations and regulatory challenges that need addressing as these technologies become more prevalent.

Overall, the articles collectively highlight the transformative potential of DL and machine learning in financial forecasting and asset management. Enhanced predictive accuracy, integration of alternative data sources, and the dynamic optimisation of portfolios are key benefits of DL methodologies identified in multiple studies.

The DL field is evolving at a fast pace. Therefore, we propose performing a systematic financial asset management review focusing on the most recent years. This review will build on the foundations laid by previous surveys, particularly those of Mirete-Ferrer et al. (2022) and address several gaps and emerging trends in the literature.

While previous reviews have extensively covered the application of DL in financial forecasting, they have yet to fully explore the latest developments and innovations in financial asset management over the past five years. Furthermore, previous reviews have highlighted the importance of integrating alternative data sources into DL models to improve predictive performance. However, practical implementation and impact issues need to be further explored. Jiang (2021) has highlighted the potential of these data sources to enhance the predictive performance of DL models. Understanding what has been done will help to identify new opportunities to leverage various data types in DL models.

Moreover, previous reviews have not considered carbon and electricity markets in their analysis of DL applications in financial asset management. This exclusion is noteworthy because these markets present unique characteristics and challenges that could benefit significantly from advanced predictive models. Including these markets in this survey provides a more comprehensive understanding of how DL models can be tailored to address diverse financial markets' specific needs and challenges, thus enhancing their applicability and impact.

In summary, a systematic review of recent empirical financial asset management articles in this field over recent years (2018 to 2023) is justified given:

- **Advancements in predictive accuracy:** Given the rapid evolution of the DL field, a systematic review focusing on the latest developments of DL in financial asset management and financial markets is essential to build on previous surveys and address emerging trends and gaps in the literature.
- **Integration of alternative data sources:** A more recent survey will help understand the advancements and opportunities in leveraging diverse data types to enhance DL models.
- **Inclusion of Carbon and Electricity Markets:** Previous reviews have not considered carbon and electricity markets, which present unique characteristics and challenges. Including these asset categories could significantly benefit general predictability.
- **Understanding best practices**: By systematically reviewing the most recent literature, we can identify best practices in implementing DL models, such as baseline models and robust performance metrics.

This review summarises recent empirical findings and identifies future research gaps and directions, thereby contributing to developing more effective and sophisticated DL models.

Our work is structured as follows. We first provide the theoretical background in Section 2. Section 3 explains the methodological approach. Section 4 analyses the results. Section 5 provides a robustness analysis. Finally, we conclude in Section 0.

## 2 Theoretical background

This section provides a brief theoretical background in financial asset management and DL. While not exhaustive, we will describe the literature's main topics, data sources, performance metrics and DL models.

### 2.1 Topic

Our work will follow a similar approach to that of Mirete-Ferrer et al. (2022). We divide the *portfolio construction* task into value/factor investing, algorithmic trading, portfolio management, and price forecasting.

#### 2.1.1 Value/factor investing

Asset pricing is a core area of finance that involves determining the fair value of financial assets. The goal is to understand how assets are priced in the market, considering risk, return, and various factors that affect financial asset market performance. Many asset managers use these models to capture excess returns and build their portfolios around these factors. Ross' (2013) Arbitrage Pricing Theory (APT) significantly developed this field. Unlike the Capital Asset Pricing Model (CAPM) (Lintner, 1975; Sharpe, 1964). APT considers multiple factors that might affect asset returns. Several authors proposed multi-factor models that are not solely based on market risk. Examples are Fama & French (1993, 2015, 2018), Carhart (1997), and Hou et al. (2021).

Meanwhile, DL factor models outperform the traditional linear models, implying that the relationship between the stock market excess returns and the factors is non-linear rather than linear (Nakagawa et al., 2018). Moreover, they outperform non-linear traditional models as no prior factor specification is needed. However, due to their *black-box* nature, DL models need more interpretability to be more useful in asset pricing. These baseline models should be portfolios built on traditional models, i.e., Fama and French three- and five-factor models (Fama & French, 1993, 2015), as these are the most recognised in the financial literature and by practitioners.

#### 2.1.2 Algorithmic trading

Algorithmic trading refers to using computer algorithms to automate the buying and selling financial securities (Ozbayoglu et al., 2020). These algorithms are designed to decide the timing, pricing and quantity of trades based on predefined criteria, including market conditions and historical data. Algorithmic trading aims to maximise efficiency and profitability by executing trades at speeds and frequencies that would be impossible for human traders (see Huotari et al., 2020). This approach has revolutionised financial markets, making trading faster and often more cost-effective.

This technological advancement has profoundly impacted portfolio management by enabling dynamic adjustments and real-time optimisation of investment portfolios. As a result, there has been a significant increase in the number of quantitative analysts in the job market. Quantitative analysts specialise in DL strategies, which differ from traditional methods because they can process vast datasets and identify complex market patterns. While traditional quantitative analysts may rely on established financial models and historical trends, modern quantitative analysts leverage cutting-edge technologies to adapt to rapidly changing market conditions.

#### 2.1.3 Portfolio management

*Portfolio selection and portfolio optimisation* are often used interchangeably in finance, yet they refer to distinct processes with different focuses. Portfolio selection is the broader process of choosing a mix of assets to include in an investment portfolio. This involves identifying potential investments that align with the investor's risk tolerance, objectives, and time horizon. In contrast, portfolio optimisation refers to the quantitative process of determining the ideal asset allocation within the chosen portfolio, aiming to maximise returns for a given level of risk or to minimise risk for a specified expected return. Despite their differences, both processes are integral to portfolio management. Portfolio management encompasses creating and maintaining an investment portfolio to meet specific financial goals, with selection and optimisation forming crucial components of this holistic approach.

Diversification is a key element in portfolio management, as highlighted by Markowitz (1952). Closely linked to diversification is the concept of asset correlation. According to Goetzmann & Kumar (2008), while investors recognise the benefits of diversification, they often build portfolios without sufficiently considering asset correlations. This oversight is a primary reason why many advanced portfolio optimisation techniques perform well in-sample but tend to underperform out-of-sample.

The broader debate on active versus passive portfolio management strategies is closely connected to this issue. Active portfolio management involves selecting financial assets based on fundamental and technical analysis. This discussion ties into deep learning (DL) models, which assist in portfolio selection and optimisation decision-making.

However, some authors consider predicting future returns, volatilities, and correlations of financial assets unfeasible. For instance, DeMiguel et al. (2009) show that an equal-weighted (EW) allocation, where each asset is assigned the same weight, outperforms many commonly used portfolio optimisation methods. Other authors argue that the best portfolio allocation is based on market capitalisation weights (MW), as suggested by Sharpe (1964) and Lintner (1975).

These strategies fall under passive portfolio management, often linked to index and exchange-traded funds, which prioritise low fees and diversification. The financial world has undergone a remarkable transformation over recent decades, with more investors gravitating towards passive management strategies (Anadu et al., 2020), largely due to the high transaction costs associated with active management (French, 2008).

In academic literature, articles on portfolio management and algorithmic trading are encouraged to incorporate the naïve Buy-and-Hold (B&H) method, which can be implemented using either an EW or MW approach.

### 2.1.4    Price forecast

Price forecasting refers to the process of predicting future prices of financial assets. This practice is fundamental in finance, as it helps investors, traders, and financial institutions make informed decisions regarding buying, selling, or holding financial assets. Price forecasting is intrinsically linked to Fama's Efficient Market Hypothesis (EMH).

According to Fama (1970), financial markets are *efficient* because asset prices fully reflect all available information at any given time. EMH is divided into three forms: *weak*, *semi-strong*, and *strong*. The *weak* form asserts that current prices reflect all past trading information, implying that technical analysis cannot consistently predict future price movements. The *semi-strong* form states that prices adjust rapidly to new public information, rendering fundamental analysis ineffective or costly for gaining a nontrivial advantage. The *strong* form posits that prices reflect all public and private information, suggesting that even insider information cannot be used to achieve consistently higher returns.

Under the EMH framework, the effectiveness of price forecasting is called into question. If markets are truly efficient, as Fama argues, then price movements are largely unpredictable and follow a random walk, meaning that future price changes are independent of past price movements (Fama, 1965). This suggests that neither technical nor fundamental analysis can reliably forecast future prices to achieve abnormal returns, invalidating active portfolio management. Therefore, the main baseline for studies on this topic should be a random walk approach, i.e., assuming today's price is the best predictor for the next day of a given asset. As evidenced in Section 4, this is the main topic addressed by the authors using DL models, challenging Fama's (1970) assumptions.

### 2.2    DL models

In the DL field, there are vast possibilities for models. To understand their evolution in financial asset management, we identify the most common models and explain their main benefits in this section.

In their basic form, Artificial neural networks (ANN) are models inspired by biological neural networks, and the neuron in an ANN consists of an aggregation function that calculates the sum of the inputs and an activation function that generates the output that can capture a non-linear relationship between financial data. In this survey, we will follow the Jiang (2021) methodology. Therefore, we categorise the following models into the ANNs because they share a similar structure: backpropagation neural network, multilayer perceptron, extreme learning machines, and comparable models where the parameters of hidden nodes need not be tuned.

Although similar in structure, the autoencoders (AE) (Hinton & Salakhutdinov, 2006) are considered separately from ANNs. AEs are a type of neural network used in DL that learns efficient and compact representations (encodings) of the input data, typically for dimensionality reduction or feature extraction. They work by compressing the input into a lower-dimensional code and then reconstructing the output from this representation. The architecture of an autoencoder is designed as a sandwich of layers: The first part (encoder) reduces the input dimensions and captures the key characteristics in a compressed format, and the second part (decoder) attempts to reconstruct the original input from this compressed code. Due to its feature extraction ability, autoencoders are useful in various applications, mainly asset pricing.

Another model type is the convolutional neural networks (CNNs) (LeCun et al., 1998). CNNs are a specialised neural network used primarily to process data with a grid-like topology, such as images. CNNs excel in capturing spatial hierarchies in data by applying convolutional filters that process data through sliding windows of weights across the input, capturing features like edges, textures, and shapes. This process allows CNNs to build complex representations of images as they move through layers, where earlier layers capture basic features, and deeper layers interpret more complicated structures. Unlike traditional neural networks that require input to be flattened into vectors, CNNs maintain the spatial structure of the data, making them highly efficient for tasks such as image and video recognition, image classification, and medical image analysis. Despite its success in image and video analysis, it is extensively used within financial asset management to denoise market information.

Generative adversarial networks (GANs) (Goodfellow et al., 2014) are another model capable of denoising market information. By training a GAN with historical market data, the *generator* can produce new synthetic asset price movements that mimic real market dynamics. At the same time, the *discriminator* works to distinguish between these synthetic data points and actual historical data. This iterative process enhances the model's ability to reflect true market behaviours, providing portfolio managers with a robust tool for risk management and strategy optimisation. Moreover, GAN-generated data can help fill gaps for rare events not well represented in historical data, such as extreme market downturns, thereby allowing managers to better prepare for potential crises. Therefore, GANs have significant potential to improve decision-making processes in portfolio management through advanced, realistic simulations of financial time series.

Another model type is graph neural networks (GNNs) (Scarselli et al., 2008). They are equipped to manage complex data structures such as those in financial asset management, particularly in portfolio management, where relationships between assets significantly influence investment decisions. In portfolio management, GNNs can analyse the interconnectedness of various assets within a portfolio by treating these relationships as a graph where assets are nodes and their correlations are edges. This approach allows GNNs to capture the individual asset characteristics and their dynamic interactions, offering a comprehensive view crucial for risk assessment and optimisation. Using GNNs, portfolio managers can improve their decision-making processes with insights derived from their portfolios' deep and intricate connections, potentially leading to more robust and optimised asset allocation. This capability is especially valuable in managing portfolios where the temporal dynamics and correlations of assets, often captured in time series data, play critical roles in predicting future asset behaviour and adjusting portfolios accordingly.

Since financial data are normally in time series, recurrent neural networks (RNNs) (Elman, 1990) are the standard approach as they excel at processing sequential data. RNNs are distinguished from other neural networks by their internal memory, which captures information about what has been processed so far, effectively giving them a form of short-term memory. This is achieved through loops within the network architecture that allow information to persist across time steps. This foundational approach has enabled RNNs to learn sequences and their long dependencies, although with limitations such as the vanishing gradient problem, which led to the development of long short-term memory (LSTM) (Hochreiter, 1997). LSTMs maintain a more complex internal state to remember information for longer periods, using structures called gates, specifically, input, forget, and output gates. These gates control the flow of information: they decide what to retain in memory, what to discard, and what to output. This gating mechanism allows LSTMs to capture long-term dependencies in data, making them highly effective for applications in time series forecasting, where understanding context over time is crucial. However, a critical limitation of RNNs and LSTMs is that they have difficulty processing sequences much longer than their internal state, as the internal state can only summarise information from a limited number of time steps.

Additionally, RNNs and LSTMs are sequential, meaning they must process elements in a sequence one at a time (Lipton et al., 2015), leading to high computational costs when processing long and complex data sequences. Gated recurrent units (GRUs) (Cho, 2014) were created to overcome this issue. GRUs achieve this through a more streamlined design incorporating two gates: an update gate and a reset gate. The update gate helps the model decide how much past information (from previous time steps) needs to be passed along to the future, effectively blending the old and new information. The reset gate determines how much of the past information to forget, which helps the model to adapt more to changes in the data sequence. GRUs and LSTMs can be enhanced by introducing bidirectional learning, allowing the model to capture past and future contexts in sequence data by processing it forward and backwards.

More recently, the Transformers represent a significant advancement in the field of DL, particularly for tasks that involve processing sequential data. Introduced by Vaswani et al. (2017), the transformer model steers from traditional recurrence-based approaches and instead utilises a structure based entirely on attention mechanisms. The core idea of transformers is the attention mechanism, which allows models to weigh the importance of different parts of the input data differently. This is particularly useful in understanding the context within which elements appear, improving the model's ability to discern relationships and dependencies in data, regardless of their distance from the input sequence. This enhances the ability to handle long-range dependencies and makes them well-suited for modelling the entire sequence of historical prices or economic indicators, thereby capturing complex patterns that affect asset prices. This capability can significantly improve forecast accuracy, risk assessment, and identification of investment opportunities by providing a nuanced understanding of market dynamics. The attention mechanism is the basis for generative artificial intelligence, mainly for large language models like BERT (Devlin et al., 2018).

Deep reinforcement learning (DRL) is the main DL model in algorithmic trading and portfolio management. These models combine other DL models (e.g. LSTM predictions) with reinforcement learning principles to address complex decision-making tasks that involve sequential decision-making. DRL consists of an agent that learns to make decisions by interacting with an environment and receiving rewards or penalties based on the actions it takes. Integrating DL into this framework allows the agent to interpret high-dimensional inputs and learn optimal strategies from complex, unstructured data environments.

Among the vast number of models, Deep Q-networks (DQNs) (Mnih et al., 2015) stand out within the field of DRL. They offer substantial benefits by scaling Q-learning to manage complex, high-dimensional state spaces typical in financial markets. The core innovation in DQNs is their use of deep convolutional neural networks to approximate the Q function. This function predicts the reward for taking specific actions based on current market conditions, guiding decision-making in trading and investment strategies. DQNs process raw market data to output action values, helping select optimal actions to maximise potential rewards. This capability enables DQNs to automate trading strategies, optimise portfolio rebalancing, and manage risk effectively by adapting to real-time changes in the market. By continuously updating their policies based on new data, DQNs help portfolio managers improve consistency in performance by navigating varying market conditions and maintaining a balanced approach between risk and return by implementing risk-return ratios in the objective function.

*2.3    Input Data*

This section discusses the types of data inputs commonly used to train DL models in finance. The most prevalent type is historical data, which includes past price and volume information for financial assets. Historical data comprises open, high, low, close, adjusted close prices, and trading volumes. This data can be utilised across various timeframes such as daily, weekly, monthly, or intraday prices.

Another critical data source is fundamental data, which reflects the intrinsic value of an asset. This category includes financial statements like income statements, balance sheets, cash flow statements, and key metrics such as earnings per share, price-to-earnings ratio, and dividend yield. Fundamental analysis evaluates an asset's financial health, growth prospects, and overall performance to determine its value based on discounted cash flows.

Technical data involves market indicators derived from historical price and volume data. This includes chart patterns, moving averages, oscillators (such as the relative strength index), and other technical indicators. Technical analysts use this data to predict future price movements, assuming historical price patterns and trends tend to repeat over time.

Macroeconomic data consists of broad economic indicators that influence the overall market environment. This includes information on gross domestic product, inflation, unemployment, interest rates, and other economic indicators released by government agencies and international organisations. Asset managers utilise macroeconomic data to understand the economic context in which they invest and how different industries are affected.

In recent years, alternative data sources such as Google Trends, news, tweets, and other social media have been incorporated into analytical frameworks to gain deeper insights into market sentiment and emerging trends. These sources provide timely information on economic developments, company announcements, and geopolitical events and allow for sentiment analysis, producing sentiment factors (e.g., positive, neutral, or negative) to enhance DL predictions. We have grouped these alternative data sources under *other sources* in our work.

*2.4 Performance metrics*

In this section, we categorise the evaluation metrics used to assess the performance of the DL models proposed in the literature. We have identified two main groups of performance metrics.

The first group pertains to the performance of portfolios. These metrics are commonly found in value/factor investing, algorithmic trading, and portfolio management. A summary of these performance metrics is provided in Table 2. While total return is the most frequently used metric in the literature, the Sharpe Ratio is among the most useful. In finance, adjusting returns for the level of risk taken is crucial, typically quantified by the standard deviation. Although the Sharpe Ratio assumes normally distributed excess returns, it can be supplemented by analysing conditional value at risk (cVaR) and maximum drawdown (MDD) to better understand the downside risk or the strategy implemented.

*Table 2 - Summary of performance metrics for portfolios. Where $V_t$ is the portfolio value at time t, $V_0$ is the initial portfolio value, $\sigma_p$ is the standard deviation of the portfolio, n is the number of observations, $R_i$ is the portfolio return on a given period, $\bar{R}$ is the portfolio mean return, $\sigma_{d,p}$ is the standard deviation of the negative portfolio returns (downside deviation), $z_\alpha$ is the z-score corresponding to the confidence level α, and $VaR_p$ is the Value at Risk at the confidence level p.*

| Name | Formula | Notes |
|---|---|---|
| Portfolio-accumulated returns | $Accumulated\ Return = \frac{V_t - V_0}{V_0}$ | It represents the overall increase in value, considering all capital gains and income (dividends and interest). |
| Compound Annual Growth Rate (CAGR) | $CAGR = \left(\frac{V_t}{V_0}\right)^{\frac{1}{n}} - 1$ | represents an investment's mean annual growth rate over a period longer than one year. Smooth out the effects of volatility. |
| Standard Deviation of Returns | $the\ \sigma_p = \sqrt{\frac{1}{n-1}\sum_{i=1}^{n}(R_i - \bar{R})^2}$ | Quantifies the dispersion of individual returns around the mean return, indicating how much the returns can deviate from the expected value, measuring its risk. |
| Sharpe ratio (SR) | $SR = \frac{\bar{R} - r_f}{\sigma_p}$ | SR measures the risk-adjusted return of an investment portfolio. It indicates how much excess return is generated per unit of risk. |
| Sortino Ratio | $Sortino\ R = \frac{\bar{R} - r_f}{\sigma_{d,p}}$ | The Sortino ratio is a variation of the SR, focusing on downside risk by considering only negative deviations of returns (semi-variance). |
| Value at Risk (VaR) | $VaR_\alpha = \bar{R} + z_\alpha \sigma_p$ | It provides a probabilistic assessment of the maximum expected loss for a given confidence level. |
| Conditional Value at Risk (cVaR) | $cVaR_\alpha = \frac{1}{1-\alpha}\int_\alpha^1 VaR_p\ dp$ | Provides a more comprehensive risk assessment by focusing on the end of the loss distribution. |
| Maximum drawdown (MDD) | $MDD = \frac{Through\ Value - Peak\ Value}{Peak\ value}$ | It is a measure of downside risk that highlights the worst loss experienced. |

The second group encompasses the performance metrics used in price forecasting. We identify three categories within price forecasting: regression, categorical, and probabilistic forecasting. Most articles employ regression forecasting, which involves predicting a specific numerical value for a future observation of a time series or a dependent variable. This can include estimating the future returns of a given asset. Table 3 summarises the most used metrics for regression forecasting. Among these, RMSE (Root Mean Squared Error), MAE (Mean Absolute

Error), and MAPE (Mean Absolute Percentage Error) are frequently utilised. However, we would like to highlight the importance of Theil's U2 statistic (Theil, 1965), which is extremely useful for testing Fama's (1970) market efficiency when a naïve model is not provided. The Diebold-Mariano (DM) test (Diebold & Mariano, 2002) is valuable when comparing two or more competing forecast models.

*Table 3 - Summary of performance metrics for regression forecasting models. In these formulas, $y_i$ denotes the observed values, $\hat{y}_i$ represents the predicted values from the model, $\bar{y}$ is the mean of the observed values, $n$ is the number of observations, $d_t$ is the loss differential series, where $d_t = L(e_{1t}) - L(e_{2t})$, $L$ is the loss function, $e_{1t}$ and $e_{2t}$ are the forecast errors from two models, $\hat{f}_d(0)$ is an estimate of the spectral density of the loss differential at frequency zero, and $T$ is the number of forecasts.*

| Name | Formula | Notes |
|---|---|---|
| R-squared (R²) | $R^2 = 1 - \frac{\sum_{i=1}^{n}(y_i - \hat{y}_i)^2}{\sum_{i=1}^{n}(y_i - \bar{y})^2}$ | Indicates the model's goodness of fit, showing how well the independent variables explain the variability of the dependent variable. It is the main performance metric for asset pricing. |
| Root Mean Squared Error (RMSE) | $RMSE = \sqrt{\frac{1}{n}\sum_{i=1}^{n}(y_i - \hat{y}_i)^2}$ | Provides an estimate of the standard deviation of the forecast errors. It is sensitive to outliers as it squares the errors, giving more weight to larger errors. |
| Mean Absolute Error (MAE) | $MAE = \frac{1}{n}\sum_{i=1}^{n}|y_i - \hat{y}_i|$ | MAE is less sensitive to outliers compared to RMSE. |
| Mean Absolute Percentage Error (MAPE) | $MAPE = \frac{100\%}{n}\sum_{i=1}^{n}\left|\frac{y_i - \hat{y}_i}{y_i}\right|$ | MAPE provides a percentage error, which is easier to interpret and compare across different datasets. |
| Theil's U2 | $Theil's\ U2 = \frac{\sqrt{\frac{1}{n}\sum_{i=1}^{n}(y_i - \hat{y}_i)^2}}{\sqrt{\frac{1}{n}\sum_{i=1}^{n}(y_i - y_{i-1})^2}}$ | Theil's U2 value of less than 1 indicates that the model has better predictive power than the naive model. |
| Diebold-Mariano (DM) test | $DM = \frac{\bar{d}}{\sqrt{\frac{2\pi \hat{f}_d(0)}{T}}}$ | The DM test compares the predictive accuracy of two competing forecast models. |

Some studies prefer categorical forecasting, which estimates the direction of price movement—whether it will go up, down, or remain stable. This approach is used when the outcome of interest is qualitative rather than quantitative, focusing on classifying future events into predefined categories. Table 4 provides a summary of the performance metrics used in categorical forecasting. These metrics are widely used in the literature adopting this methodology.

*Table 4 - Summary of performance metrics for classification forecasting models. In these formulas, TN is the number of true negatives, FN is the number of false negatives, TP is the number of true positives, and FP is the number of false positives.*

| Name | Formula | Notes |
|---|---|---|
| Accuracy | $Accuracy = \frac{TP + TN}{TP + TN + FP + FN}$ | Accuracy measures the proportion of true results (both true positives and true negatives) among the total number of cases examined. |
| Precision | $Precision = \frac{TP}{TP + FP}$ | Also known as positive predictive value, it is the ratio of correctly predicted positive observations to the total predicted positives. |
| Recall | $Recall = \frac{TP}{TP + FN}$ | Also known as the sensitivity or true positive rate, it is the ratio of correctly predicted positive observations to all observations in the actual class. |

| F1 score | $F1\ Score = 2 \times \dfrac{Precision \times Recall}{Precision + Recall}$ | The F1 score is the harmonic mean of precision and recall. |

Furthermore, another approach involves forecasting a confidence interval for the predictions. This method provides a better understanding of the model's performance in predicting points. Table 5 summarises these performance metrics commonly found in the literature using this methodology.

*Table 5 - Summary of performance metrics for interval forecasting models. Where $U_i$ is the upper bound, $L_i$ is the lower bound of the prediction interval for observation i. $I(y_i \in [L_i, U_i])$: Indicator function that equals 1 if $y_i$ is within $[L_i, U_i]$, otherwise 0.*

| Name | Formula | Notes |
| --- | --- | --- |
| Forecast Interval Accuracy Measure (FINAW) | $FINAW = \dfrac{1}{n}\sum_{i=1}^{n}\left(\dfrac{U_i - L_i}{2} + \dfrac{\left\|y_i - \dfrac{U_i + L_i}{2}\right\|}{\dfrac{U_i - L_i}{2}}\right)$ | Lower FINAW values indicate better performance, implying that the intervals are narrow and the actual values are close to the interval centres. |
| Average Width of Prediction Intervals (AWD) | $AWD = \dfrac{1}{n}\sum_{i=1}^{n}(U_i - L_i)$ | It provides a simple assessment of the precision of the intervals, with narrower widths indicating more precise predictions. |
| Forecast Interval Coverage Probability (FICP) | $FICP = \dfrac{1}{n}\sum_{i=1}^{n} I(y_i \in [L_i, U_i])$ | It evaluates the reliability of the prediction intervals, ideally aligning with the nominal confidence level (e.g., 95%). |

By categorising these evaluation metrics, we provide a comprehensive overview of the methods used to assess the performance of DL models in finance, highlighting both their applications and limitations. This structured approach helps understand how different metrics can be applied to evaluate various aspects of DL models, from portfolio performance to price forecasting accuracy.

## 3 Methodology

Using a standard methodology to conduct a review not only supports the quality of the review but also allows researchers to replicate the review study. Given this, the study adopts the PRISMA standard for conducting the review process. PRISMA stands for Preferred Reporting Items for Systematic Reviews and Meta-Analyses. Since the study examines a single database, few adjustments were required, as explained in the following stages.

### 3.1 Identification

We identified the current work trends, research gaps, and keywords most prominent in machine learning. Using the Scopus[1] database for research material provided a comprehensive and relevant collection of articles. While the review is limited to English-language articles indexed in Scopus, this decision was made to ensure consistent quality and peer-reviewed reliability.

The search strategy was adopted to filter the articles based on keywords in the title and abstract. We searched for "deep learning" or "neural networks" in combination with "asset management," "portfolio management," "portfolio optimisation," "asset pricing," "price forecasting," or "algorithmic trading." The search was restricted to empirical articles published in English between 2018 and 2023. This strategy was designed to focus on recent advancements and the most relevant studies in the field. Articles without a valid DOI or not published in peer-reviewed journals were excluded from the search results.

The initial search yielded 934 articles. These articles underwent a two-stage screening process to ensure relevance and quality. These articles were screened further to ensure their relevance and quality, providing a robust foundation for the systematic review. This approach enables a focused and thorough examination of the advancements and applications of DL in financial asset management, highlighting current trends and identifying areas for future research.

---

[1] Accessed on the 21st of February 2024.

*3.2 Screening*

In the second stage of our systematic review, we meticulously screened all 934 articles identified during the initial search to ensure their relevance to financial asset management. This process was critical to maintaining the focus and quality of the review. Each article underwent an abstract and full-text screening to evaluate its alignment with the predefined inclusion criteria. Three hundred twenty-one articles were excluded during this stage for various reasons, detailed below.

The most common reason for abstract exclusion was that many articles pertained to physical asset management, such as infrastructure, roads, or buildings, rather than financial assets. While valuable in their domains, these studies fell outside the scope of this review, which focuses exclusively on financial asset management. The remaining articles were subjected to full-text analysis.

During the full-text analysis some articles were eliminated because they only referenced deep learning superficially without incorporating any practical or methodological application of these models. Such papers were deemed insufficiently relevant to contribute to the understanding or advancement of deep learning techniques in financial asset management.

The reasons for these exclusions and other minor factors are systematically summarised in Table 6. After applying these rigorous criteria, 612 articles were retained for further analysis. These selected articles form the basis of this survey, providing a comprehensive overview of deep learning applications in financial asset management and ensuring that the review delivers meaningful and actionable insights within its defined scope.

*Table 6 - Exclusion criteria for the current survey.*

|  | Reason | References |
|---|---|---|
| Preselection |  | 934 |
| Rejected | Not financial asset management | 192 |
|  | Not DL | 63 |
|  | No DOI | 33 |
|  | Not empirical. | 15 |
|  | Review article | 15 |
|  | Retracted | 3 |
|  | Not in English | 1 |
| **Total** |  | **612** |

To provide a comprehensive overview of the models used in financial asset management, we did not filter by Journal. Although articles with lower quality will be included in our analysis, this will help consolidate the best practices and areas for improvement from researchers. The main journals where these articles have been published are predominantly computer science. The publications by the Journal are listed in Table 7.

*Table 7 - Number of publications by Journal.*

| Journal | No. Articles |
|---|---|
| IEEE Access | 37 |
| Expert Systems with Applications | 36 |
| Energies | 27 |
| Applied Energy | 20 |
| Mathematics | 14 |
| Neural Computing and Applications | 14 |
| Applied Soft Computing | 13 |
| Resources Policy | 13 |
| Computational Economics | 11 |
| Energy | 11 |
| Applied Intelligence | 10 |
| Applied Sciences | 10 |
| Electric Power Systems Research | 9 |
| Energy Economics | 9 |
| International Journal of Forecasting | 7 |
| Quantitative Finance | 7 |
| Soft Computing | 7 |
| Sustainability | 7 |
| Others | 350 |
| **Grand Total** | **612** |

## 3.3 Inclusion

The final stage includes creating a database for qualitative and quantitative analysis. The current study comprises 612 articles, all analysed to create the database. Contents of the selected articles were classified based on their financial domains and were systematically arranged (See

Table 8). Details of the article's title, name and number of the author(s), year of publication, name of the publishing journal, author-specified keywords, DL models employed, performance metrics, validation methods, and other relevant information of each financial application were obtained.

*Table 8 - Number of Articles per topic.*

| Topic | No. Articles |
|---|---|
| Value/factor investing | 17 |
| Algorithmic trading | 48 |
| Portfolio management | 98 |
| Price forecast | 449 |
| **Total** | **612** |

## 4 Survey findings

This section will critically analyse the results of our study. We will start by giving an overview of all articles' results. Then, we will narrow down on each topic. Due to space limitations, we will only provide the general summary statistics indicating the current state of the DL for finance research. Please note that besides the main model, we considered all DL models used within the article in our analysis. Hybrid models, such as CNN+LSTM, are both CNN and LSTM, as they can be used for different reasons.

Overall, publications using DL models in financial asset management have increased, with price forecast as the predominant topic (see Figure 1). Although algorithmic trading has had fewer publications in recent years, asset pricing has recently received greater attention from authors, mainly due to the increase in explainable artificial intelligence in time series data.

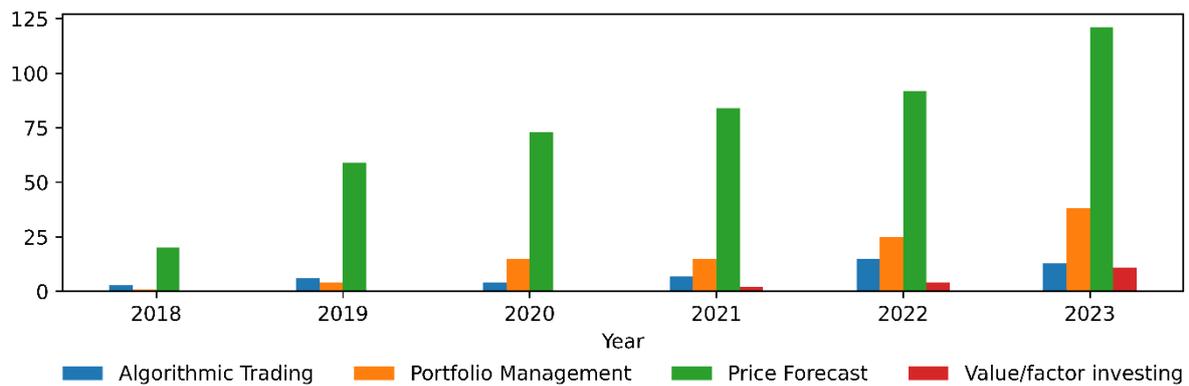

*Figure 1 - Topic publications by year.*

Regarding the type of financial asset, we have identified stocks, bonds, foreign exchange, commodities, cryptocurrencies, derivatives, and multi-asset portfolios. Although mentioned in other surveys, real estate investment funds were found among other financial assets; therefore, they were included in multi-asset portfolios. In Figure 2, equities are still the predominant financial asset studied, followed by commodities. We have created a separate analysis for electricity price forecasts, as they represent 43% of the total commodities articles. Since many practitioners use a wide range of financial assets, these results further reinforce Cremers et al. (2019) concern, as the number of articles analysing multi-asset portfolios is scarce.

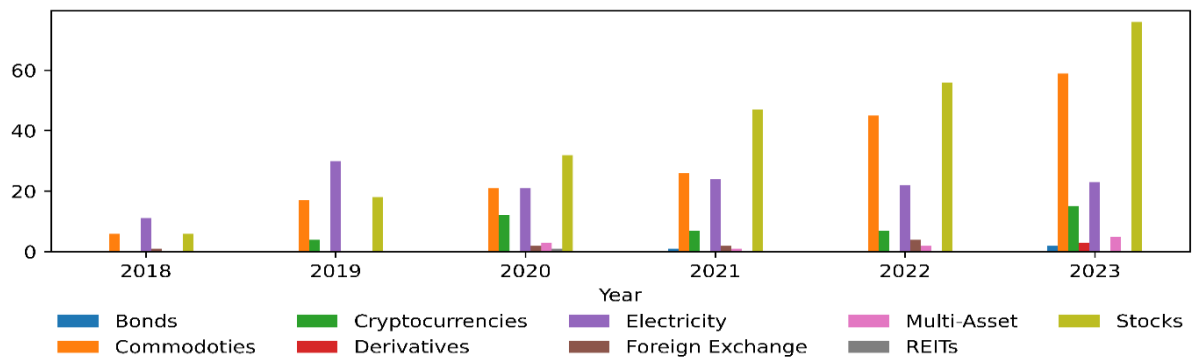

*Figure 2 - Financial asset publications per year.*

The next subsections will focus on each topic separately.

*4.1 Value/factor investing*

This topic has the lowest number of publications since 2018. Table 9 provides a sample of articles on value/factor investing. While the data source is mainly Historical and Fundamental, some articles consider Macroeconomic factors (see Alaminos et al., 2023). Therefore, 76.5% of the articles use a monthly timeframe. Moreover, due to its feature extraction, AEs are only used in 29.4% of the articles.

*Table 9 - Sample of articles in value/factor investing. FF3 and FF5 are the three- and five-factor models (Fama & French, 1993 and 2015, respectively).*

| Authors | Data sources | Model | Baseline models | Metrics |
|---|---|---|---|---|
| Gu et al. (2021) | Historical and Fundamental | AE | FF3 and other models | R2, SR |
| Son & Lee (2022) | Historical and Fundamental | GNN | FF3, AE, and other models | R2 |
| Pan et al. (2023) | Historical and Fundamental | LSTM | FF5 | R2, SR, MDD |
| Lo & Singh (2023) | Historical and Fundamental | ANN | Other models | R2 |

Gu et al. (2021) contribute by introducing AE asset pricing models. Their research highlights the superiority of AEs in reducing dimensionality and capturing complex non-linear relationships in financial data, thus enhancing the accuracy of asset pricing compared to traditional linear models. This innovative approach leverages extensive conditioning information, significantly advancing asset pricing methodologies. Pan et al. (2023) apply techniques in the Chinese stock market, focusing on non-linear asset pricing. Their findings reveal that LSTM outperforms traditional linear models in predicting stock prices and capturing the intricacies of the market. Lo & Singh (2023) utilise ANN to enhance prediction accuracy. They discuss the practical applications of these models and address the challenges of model interpretability. Son & Lee (2022) introduce a GNN multi-factor asset pricing model that uses graph theory to incorporate relationships between different assets and factors. Their approach employs GNNs to model complex interactions in financial markets.

These studies collectively highlight the superiority of DL models compared to traditional methods. However, they also underscore the persistent challenges, such as model interpretability and computational demands, that need to be addressed to realise the full potential of DL in asset pricing.

*4.2 Algorithmic trading*

For the algorithmic trading topic, we analysed 48 articles. A sample of articles are shown in

Table 10. While all articles provide other models as a baseline, only 22 consider a passive portfolio (e.g. B&H). As explained in Section 5, it is advisable that these articles contain B&H as a baseline.

*Table 10 - Sample of articles on algorithmic trading investing.*

| Authors | Financial Asset | Timeframe | Data sources | Model | Baseline | Performance metrics |
|---|---|---|---|---|---|---|
| P. Liu et al. (2023) | China and U.S. stocks | Daily | Historical and Technical | DRL with CNN+BiLSTM | B&H and other DL models | Returns, SR, MDD and MDD duration |
| Ye & Schuller (2023) | U.S. stocks | 1min | Historical and Technical | DQN | B&H, LSTM and other parameters | Returns |
| Y. Huang et al. (2023) | Oil and Gas | Daily | Historical and Technical | DADE-DQN | B&H, S&H and other parameters | Returns, SR and MDD |
| Chan et al. (2022) | EUR/GBP, EUR/JPY, EUR/USD and others | Hourly | Historical and Technical | Attention+LSTM | Simple LSTM | Returns |
| Y. Zhao et al. (2022) | U.S. stocks | Daily | Historical | LSTM | B&H and Relative Value model | Returns, SR and MDD |
| Théate & Ernst (2021) | U.S., U.K. and China stocks | Daily | Historical | TDQN | B&H, S&H, Trend following and Mean Reversion | Returns, SR, Sortino R, Std, MDD, MDD duration and P&L ratio |
| Li et al. (2021) | Gold | Daily | Historical | VMD-ICSS-BiGRU | B&H and other strategies | Returns |
| S.-H. Huang et al. (2021) | U.S. stocks | 1min | Historical and Technical | DRL | B&H, Constant Proportion and other DRL models | Returns and SR |
| Li et al. (2019) | U.S. stocks | 1min | Historical and Technical | DQN with AE+LSTM | B&H, basic A3C and DQN | Returns and SR |

We found that most articles published use a DRL model (27 out of 46), of which 7 are specifically the DQN. These models are trained to make decisions, i.e., to buy, hold or sell the financial asset. However, in the remaining 19 articles, the primary use of the model is to forecast prices. Then, rules are created for trade based on those predictions. The DL models used in this topic are shown in Figure 3.

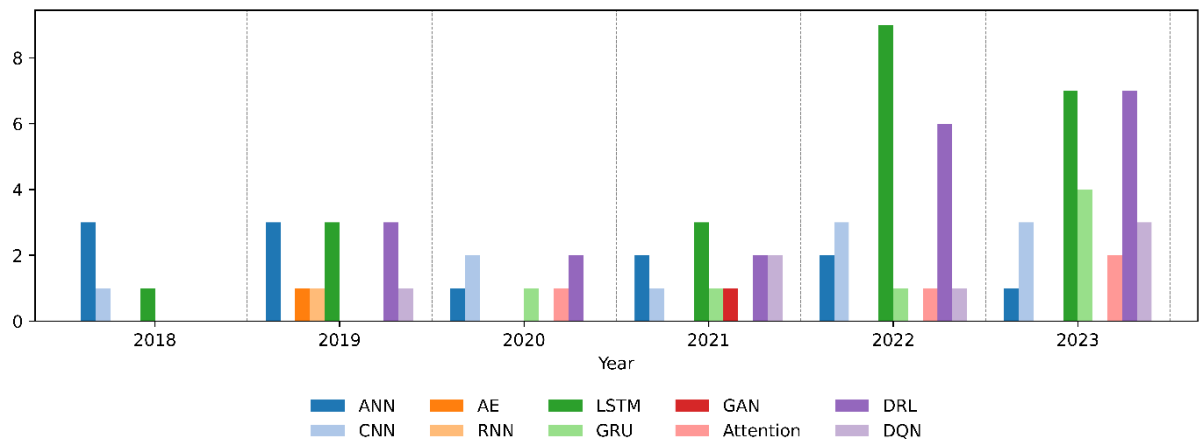

*Figure 3 - DL models used in algorithmic trading investing.*

The attention models have started to be implemented recently (see Chan et al., 2022; Lei et al., 2020). However, we found no GNN being applied to algorithmic trading. Moreover, 58.3% of the articles that use ANN are integrated with other models (see Ceffer et al., 2018; Kalariya et al., 2022).

Regarding timeframe, this is the topic with most articles using intraday data. Nonetheless, the daily timeframe is still preferred (Figure 4). For example, Ning et al. (2021) use the smallest timeframe (1 second) within the United States stock market. Due to this short timeframe, the most common data inputs are Historical and Technical, representing 52% and 37%, respectively. However, more recently, we have seen tweet volume, Google search trends, and other social media as features for the price prediction of cryptocurrencies (see Belcastro et al., 2023; Kalariya et al., 2022).

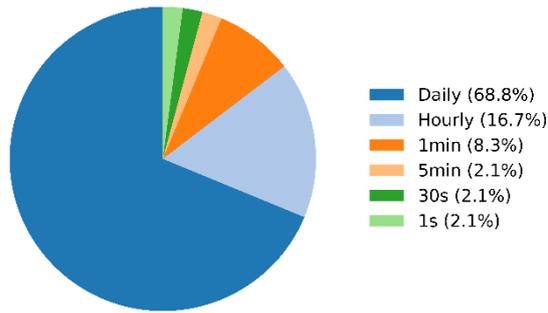

*Figure 4 - Timeframe analysis in algorithmic trading.*

Finally, returns are used in all articles as a performance metric. However, the methodology needs to improve to be more risk-sensitive. Authors should incorporate at least the Sharpe Ratio and MDD in their performance metrics, as they are only present in 30 (62.5%) and 19 (39.6%) articles, respectively.

## 4.3 Portfolio management

Portfolio management is the second most published topic in our survey. Table 11 provides a sample of articles on portfolio management topics.

*Table 11 - Sample of articles in portfolio management.*

| Authors | Financial Asset | Timeframe | Data sources | Model | Baseline | Performance metrics |
|---|---|---|---|---|---|---|
| Jang & Seong (2023) | U.S. stocks | Daily | Historical and Technical | DRL with CNN | DRL with ANN, EW and other articles | Returns, SR and MDD |
| Behera et al. (2023) | India stocks | Monthly | Historical | ANN | Other ML models | Returns, Std and VaR |
| Du (2022) | CSI 300 Index, China and U.S. stocks. | Daily | Historical, Technical and Fundamental | LSTM | Other ML models | SR, Sortino R., MDD and Turnover rate |
| Barua & Sharma (2022) | MSCI Asia Pacific sector Indices | Daily | Historical and Technical | CNN+BiLSTM | CNN, LSTM, BiLSTM and CNN-LSTM | Returns, Std, SR and Herfindahl Index |
| Aboussalah et al. (2022) | U.S. stocks | Daily | Historical | DRL with CNN | B&H (EW) and other model combinations | Returns |
| Betancourt & Chen (2021) | Bitcoin, Ethereum and Litecoin | 30min, 6h and daily | Historical | DNA-S | CNN, DQN and DRL | Returns and SR |
| Wang et al. (2020) | U.K. stocks | Daily | Historical and Technical | LSTM+MV | SVM, RAF, ANN, and ARIMA | Returns, Std, SR, cVaR, MDD and Sortino R |
| Soleymani & Paquet (2020) | U.S. stocks | Daily | Historical and Technical | DRL with CNN | B&H (MW) | SR and MDD |
| Vo et al. (2019) | U.S. stocks | Daily | Historical and Fundamental | DRL with BiLSTM | B&H (MW) | Returns, Std, SR and ESG score |
| Y. Zhao et al. (2018) | U.S. stocks | Daily | Historical | Psi Sigma Network | ANN, RNN, B&H (EW and MW) and ARIMA | Returns, SR, Sortino R., cVaR and MDD |

In portfolio selection, the main goal of the DL models is to help the asset manager select which assets will be incorporated into their investment portfolio. To do so, the DL is used to forecast prices (Behera et al., 2023), direction (Moon & Kim, 2023) or sentiment analysis (Z. Huang & Tanaka, 2022) from alternative data sources, for example, social media. An article that uses an innovative methodology was published by Adosoglou et al. (2021). The authors use a Distributed Memory Model of Paragraph Vectors mode of Doc2Vec to cluster similar companies based on financial reports. This methodology helped the portfolio selection process, beating the B&H (EW and MW) and other baseline models.

In portfolio optimisation, the main task is to decide on the weights (49%) each asset will have in the investment portfolio. Figure 5 shows the main models used to forecast the optimal weights.

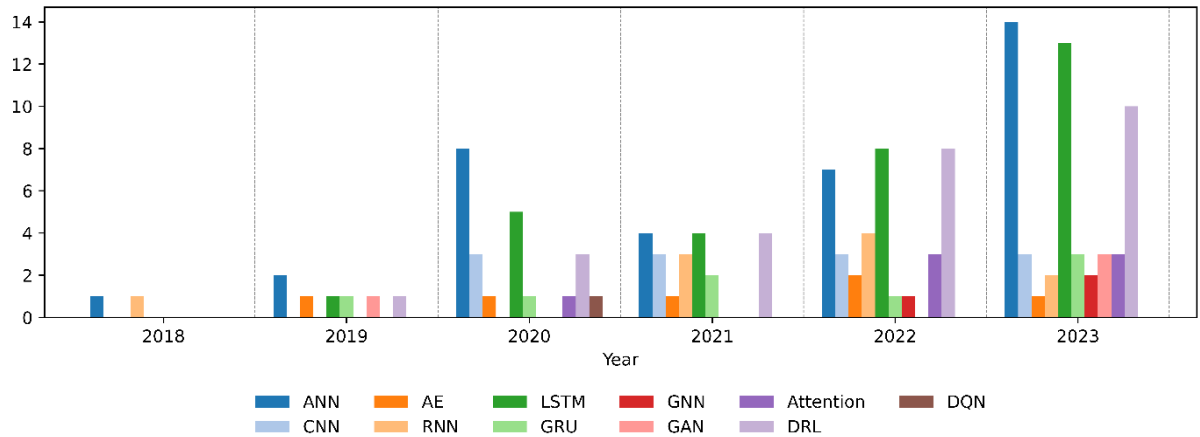

*Figure 5 - DL models used by year in portfolio management to forecast optimal weights.*

The most used models are the ANN, LSTM and DRL. ANNs are the main models in 26 out of the 36 articles where they are present (see Behera et al., 2023; Chen et al., 2020). Conclusions are similar to the algorithmic trading topic. The main models used in this topic do not include recent developments in computer science. For example, the number of articles using attention mechanisms and recent developments in DRL are still scarce.

Most studies (84.7%) use stocks in their empirical analysis. The preferred timeframe is daily and monthly, with 71 and 14 articles, respectively. Furthermore, the most used data source is the Historical prices (Figure 6). While technical analysis has recently received more attention from authors, using other input data (see Z. Huang & Tanaka, 2022) to reach the DL model's full potential would be advantageous.

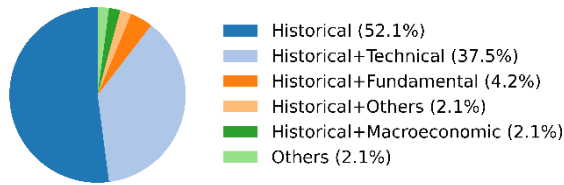

*Figure 6 - Input data used in portfolio management.*

Returns are the most used performance metric. Risk-adjusted returns, such as SR and Sortino R., are not commonly used, as they are only present in 55 and 16 articles, respectively. The most common risk measure is MDD, present in only 33 articles. Moreover, the number of articles using the baseline models B&H (EW or MW) are still scarce (36.7%).

### *4.4  Price forecast*

Researchers have explored the price forecast topic the most. They represent 73.4% of the total publications studied in this survey. Table 12 shows a sample of articles on price forecasting. However, some surveys focus on the stock market (see Jiang, 2021; Olorunnimbe & Viktor, 2023), the most financial assets studied are commodities, of which electricity represents almost half (Figure 7).

*Table 12 - Sample of articles on price forecast.*

| Article | Financial Asset | Data studied | Timeframe | Data sources | Main Model | Baseline | Horizon | Performance metrics |
|---|---|---|---|---|---|---|---|---|
| S. Zhang et al. (2023) | WTI crude oil | 02/08/2010 to 31/12/2019 | Daily | Historical | VMD-SE-GRU | ANN, GRU and LSTM | Next step | RMSE, MAE, MAPE and R2 |
| Md et al. (2023) | Samsung stock | 23/11/2016 to 23/11/2021 | Daily | Historical | Multilayer Sequential LSTM | RNN, ANN, CNN and LSTM | Next step | R2, Adj. R2, NMRSE and RMSE |
| Zaheer et al. (2023) | Shanghai Composite Index | 03/08/1997 to 24/01/2022 | Daily | Historical | CNN-RNN-LSTM | CNN, RNN, LSTM, CNN-RNN, and CNN-LSTM | Next step | MAE, RMSE and R2 |
| Jakubik et al. (2023) | Bitcoin | 12/09/2013 to 18/06/2019 | Daily | Historical and Others | BiLSTM | LSTM, CNN and Random Forest | Next step | MSE, RMSE, AUROC, Accuracy, Precision and F1 |
| Uddin et al. (2023) | Russell 3000 stocks | 01/01/1990 to 30/12/2020 | Daily | Historical, Technical, Fundamental and Macroeconomic | DY-GAP (GNN with attention) | SVM, LSTM, ANN, and other models | Next Step | RMSE, MAE, MAPE |
| Staffini (2022) | Italian stocks | 03/01/2005 to 30/04/2021 | Daily | Historical and Technical | DCGAN (with CNN+BiLSTM and CNN) | LSTM, GAN, and other models | Multi-step (1, 5) | RMSE, MAE and MAPE |
| Zhou et al. (2022) | Carbon | 14/03/2014 to 31/08/2021 | Daily | Historical | CEEMDAN-LSTM | ANN, LSTM, GRU and other models | Multi-step (1, 2, 3) | RMSE, MAE, MAPE and R2 |
| K. Zhang et al. (2022) | Coal | 1/01/2011 to 10/12/2020 | Daily | Historical | VMD-Attention-LSTM-SVR | Other variations of LSTM | Multi-step (1, 2, 3, 4) | RMSE, MAE, MAPE and R2 |
| Lin et al. (2021) | S&P 500 and CSI 300 | 10/11/2008 to 20/02/2019 | Daily | Historical | CEEMDAN-LSTM | LSTM, SVM, ANNs and other models | Next Step | MSE, MAE and R2 |
| Memarzadeh & Keynia (2021) | Electricity | 2006 and 2018 | Hourly | Historical | Wavelet-LSTM | ANN and Abedinia et al. (2017) model | Next step | MAPE, RMSE, MAE and VAR |
| Y. Huang et al. (2021) | Carbon | 01/11/2017 to 31/10/2019 | Daily | Historical | VMD-GARCH/LSTM-LSTM | LSTM and ANN variations | Multi-step (1, 2, 4, 6) | RMSE, MAE, MAPE and DM |
| Jaquart et al. (2021) | Bitcoin | 04/03/2019 to 10/12/2019 | 1min | Historical and Others | LSTM | GRU, ANN and other models | Multi-step (1, 5, 15, 60) | Accuracy |
| Livieris et al. (2020) | Gold | 01/01/2014 to 01/04/2018 | Daily | Historical | CNN-LSTM | LSTM, ANN and Support Vector Regression | Multi-step (4, 6, 9) | MAE, RMSE, Accuracy, Area under curve, Sensitivity and Specificity |
| Qiao & Yang (2020) | Electricity | 01/1997 to 12/2020 | Monthly | Historical | WT-SAE-LSTM | ANN, LSTM, BiLSTM and SAE-LSTM | Next step | RMSE, MAPE, MAE, RMSPE, U1 and U2 |
| Dutta et al. (2020) | Bitcoin | 01/01/2010 to 30/06/2019 | Daily | Historical, Technical, Fundamental and Others | GRU-Dropout-GRU | GRU, LSTM and ANN | Next step | RMSE |
| J. Cao et al. (2019) | U.S., Hong Kong, Germany and Chinese Stock Index | 13/12/2007 to 12/12/2017 | Daily | Historical | CEEMDAN-LSTM | LSTM, SVM and ANN | Next step | RMSE, MAE and MAPE |
| Wu et al. (2019) | WTI crude oil | 06/01/1986 to 06/06/2016 | Daily | Historical | EEMD-LSTM | LSTM, ANN and other models | Multi-step (1, 2, 3, 4) | RMSE, MAPE, Dstat and DM |
| Ji et al. (2019) | Bitcoin | 29/11/2011 to 31/12/2018 | Daily | Historical and Fundamental | LSTM and ANN | DNN, LSTM, CNN, ResNet, CRNN, Ensemble and Support Vector Model | Next step | MAPE |
| Y. Liu (2019) | S&P500 and Apple stock | 03/01/2000 to 29/11/2013 | Daily | Historical | LSTM | SVM and GARCH | Next Step | RMSE |
| Ugurlu et al. (2018) | Electricity | 01/01/2013 to 21/12/2016 | Hourly | Historical and Fundamental | Multilayer GRU | ANN, CNN, RNN, LSTM, GRU and naïve | Next day (24h) | MAE and DM |

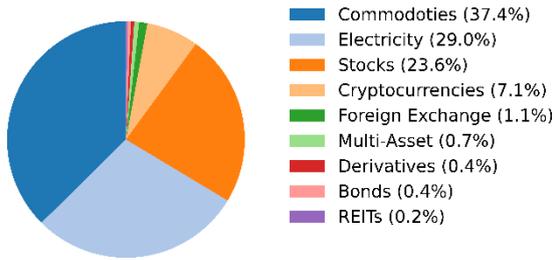

*Figure 7 - Financial asset studies in price forecast topic.*

The most common type of forecast is regression. Interval and classification forecast only accounts for 6.4% and 5.1%, respectively. Some studies perform multiple options, either regression-interval (Y. Cao et al., 2023) or regression-classification (Ji et al., 2019). One issue of concern is the low use of a naïve predictor as a baseline. They are present in only 4% of the articles, either from implementing a model or using Theil's U2 as a performance metric.

Regarding the DL models used, like other surveys (see Mirete-Ferrer et al., 2022; Nazareth & Reddy, 2023), LSTMs continue to be the most used model (Figure 8) and are gaining momentum. ANNs are present in 63.6% of the total number of articles. Also, ANNs are within the main model in 54.9% of the articles in which they appear. Incorporating attention mechanisms (K. Zhang et al., 2022) and GNNs (G. Zhao et al., 2023) has also gathered recent attention from authors, leading to better predictions.

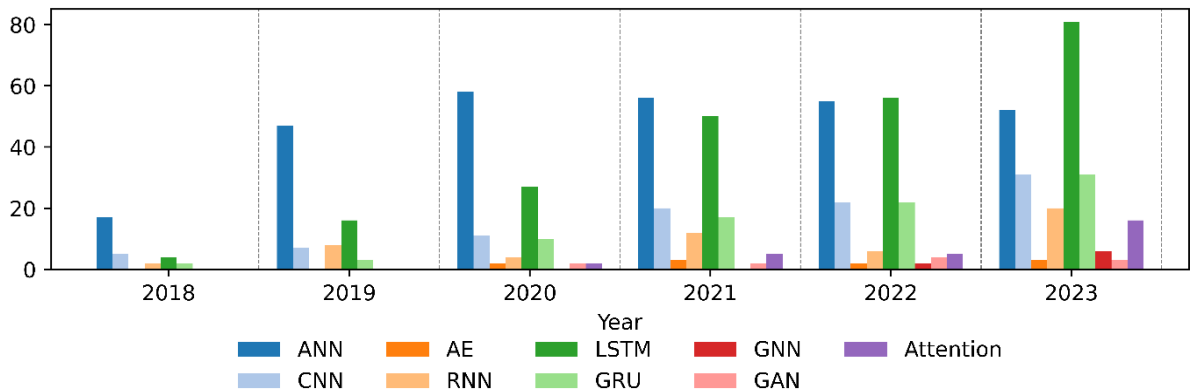

*Figure 8 - DL models in price forecast topic.*

Jiang (2021) and Nazareth & Reddy (2023) underscored the importance of integrating alternative data inputs. However, Figure 9 shows that the most common input is Historical only, accounting for 65.1% of the total publications on this topic. This is due to incorporating decomposition[2] techniques in the architecture. For example, CEEDAM[3] and VMD[4] have been recently incorporated in Zhou et al. (2022) and S. Zhang et al. (2023), respectively. Macroeconomic and other input data are still the least used.

---

[2] Decomposition refers to breaking down complex data into simpler, more manageable components to improve model performance and interpretability.
[3] Complementary Ensemble Empirical Mode Decomposition and Adaptive Multi-scale Noise Filtering.
[4] Variational Mode Decomposition.

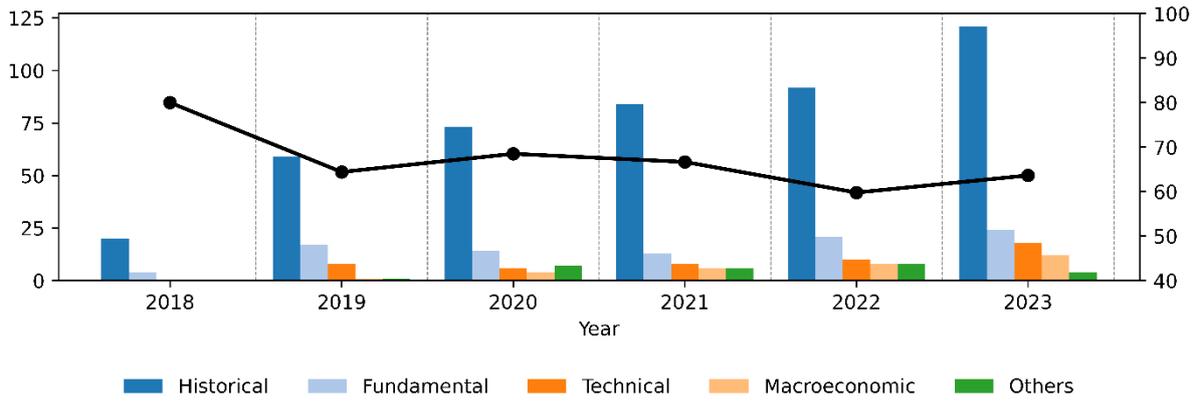

*Figure 9 - Data inputs used by year in price forecast topic (LHS). The line represents the percentage of articles using only Historical data (RHS).*

The shortest timeframe available from publicly accessible databases is normally daily data. Therefore, the most common input timeframe is daily data (Figure 10 - Timeframe analysis in the price forecast topic (Figure 10). Intraday data articles increased from 11 to 28 in 2018 and 2023, respectively.

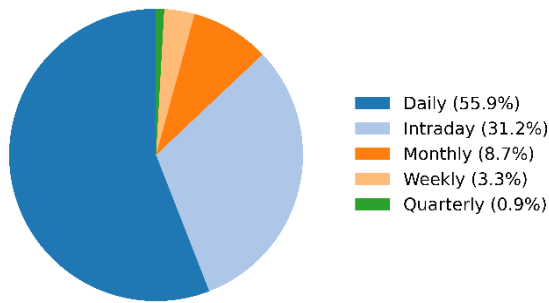

*Figure 10 - Timeframe analysis in the price forecast topic.*

Most studies focus on predicting the next step (66.6%), i.e. the next day or month, depending on the timeframe. Longer timeframes are studied in a single horizon (2.0%) or a multi-step horizon (20.3%). The number of articles using naïve forecast methods is still scarce (4.0%).

## 5 Robustness Analysis

### 5.1 Non-Electricity Price Forecast

Electricity studies differ from the other articles due to their data timeframe and horizon forecast. As they account for 29.0% of the articles on this topic, we analyse the results excluding these articles.

Analysis results are similar to those in the section 4.4. The most studied financial assets (Figure 11) are Commodities (52.7%), followed by stocks (33.2%). LSTM and ANNs are the most common models (Figure 12). The number of articles using recent models, such as attention mechanisms and GNNs, is also increasing. Historical input is the most used data input (Figure 13). Finally, using a naïve forecast baseline is also scarce (2.8%).

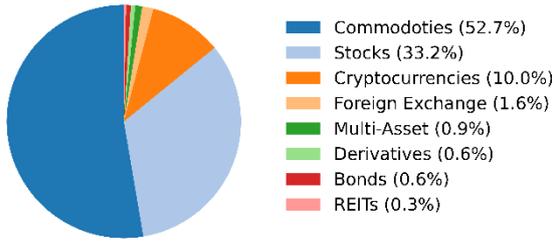

*Figure 11 - Financial asset studies on price forecasting excluding electricity.*

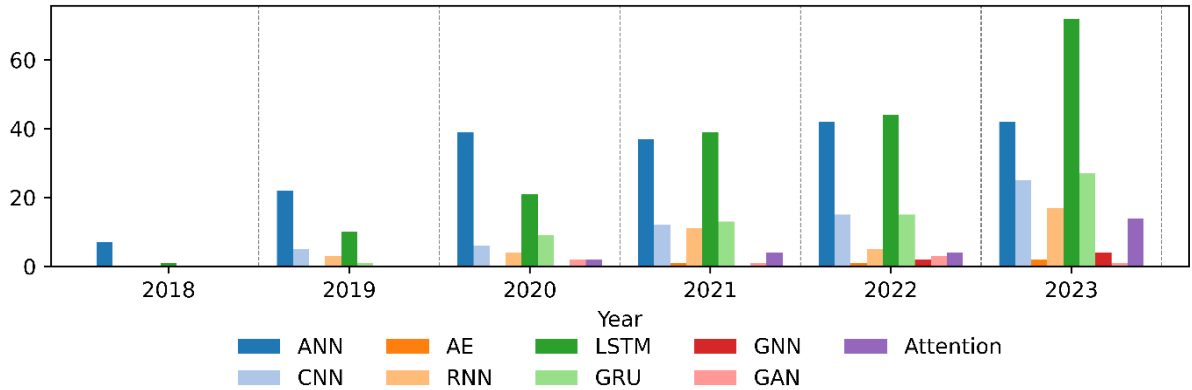

*Figure 12 - DL models in price forecast topic excluding electricity.*

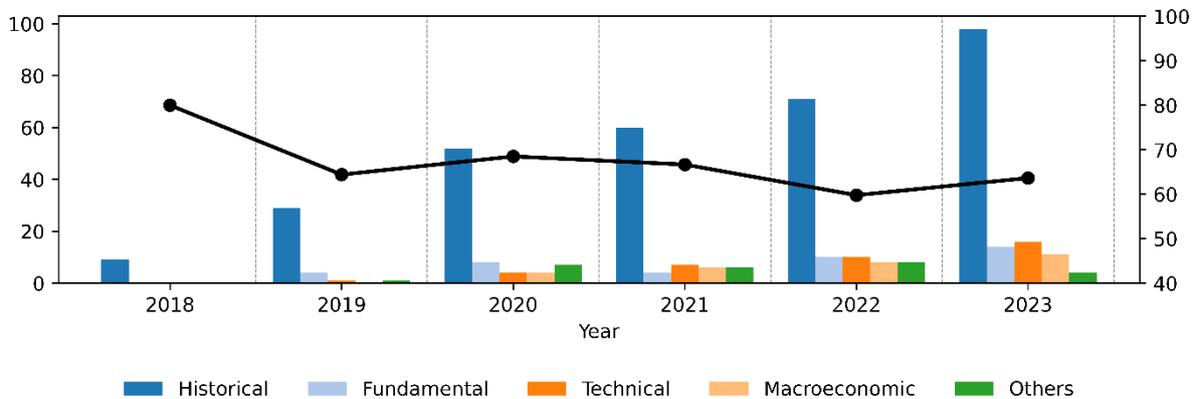

*Figure 13 - Data input used by year in price forecast topic excluding electricity (LHS). The line represents the percentage of articles using only Historical data (RHS).*

However, the results are distinct when we analyse the timeframe and horizon. The shortest timeframe available from publicly accessible databases is normally daily data for stocks and commodities. However, regarding electricity, this data is often an hourly interval (Figure 14).

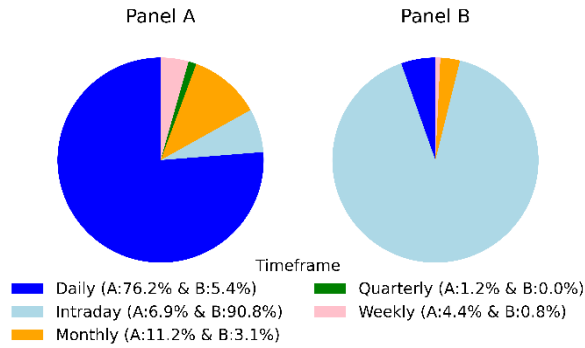

*Figure 14 - Timeframe used in price forecast. Panel A for studies without considering electricity. Panel B is for electricity articles only.*

Finally, the horizon forecast changes within these two groups (Figure 15). Most non-electricity studies focus on predicting the next step, i.e. the next day or month, depending on the timeframe. However, the predominant time frame for electricity articles is hourly (Memarzadeh & Keynia, 2021) and sometimes 30 minutes (Lu et al., 2022). Therefore, the main forecast horizon is the next day, i.e. 24 or 48 hours ahead.

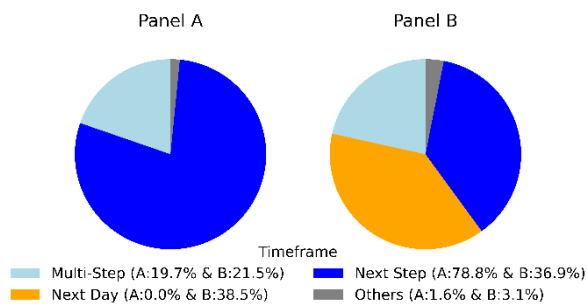

*Figure 15 - Horizon forecast for price forecast. Panel A for studies without considering electricity. Panel B is for electricity articles only. Please note that the Next Day within Panel A has been included in Next Step.*

### 5.2 Journal Inclusion Criteria

We further analyse the results of this review by incorporating a journal filter. We only included articles in the Academic Journal Guide from the Chartered Association of Business Schools 2024 guide in this section.

Table 13 and Table 14 show the number of articles per Journal and topic. Price forecasting remains the most explored topic and is gaining momentum (Figure 16). However, the percentage of articles on the price forecasting topic is significantly lower than before. It dropped from 73.37% to 58.18%, while portfolio management and value/factor investing increased their presence among the topics published by authors.

*Table 13 - Number of publications by Journal with journal filter.*

| Journal | No. Articles |
|---|---|
| Expert Systems with Applications | 36 |
| Resources Policy | 13 |
| Computational Economics | 11 |
| Energy Economics | 9 |
| Quantitative Finance | 7 |
| International Journal of Forecasting | 7 |
| Journal of Cleaner Production | 6 |
| Annals of Operations Research | 6 |
| Others | 70 |
| **Total** | **165** |

*Table 14 - Number of Articles per topic with journal filter.*

| Journal | No. Articles |
|---|---|
| Price Forecast | 96 |
| Portfolio Management | 44 |
| Algorithmic Trading | 14 |
| Value/factor investing | 11 |
| **Total** | **165** |

After applying the filter, results show similar conclusions. Stocks and commodities remain the most asset studied (Figure 17). Historical data is the most common data input (Figure 18). Finally, the most common DL model used is LSTM, with GNN and attention mechanisms gathering recent attention from authors (Figure 19).

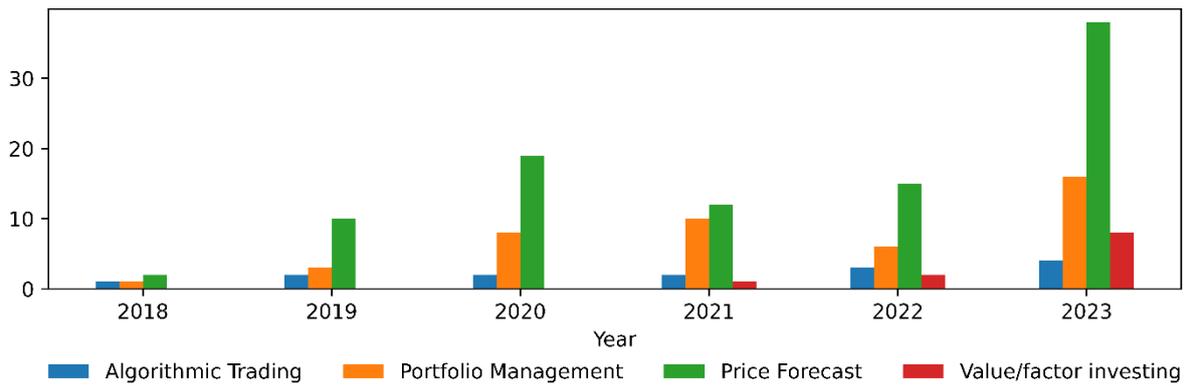

*Figure 16 - Topic publications by year with journal filter.*

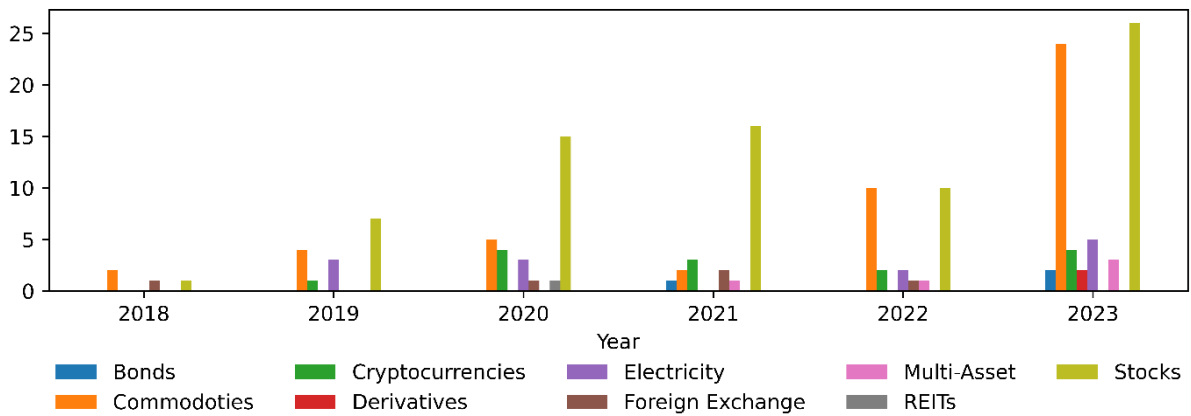

*Figure 17 - Financial asset publications per year with journal filter.*

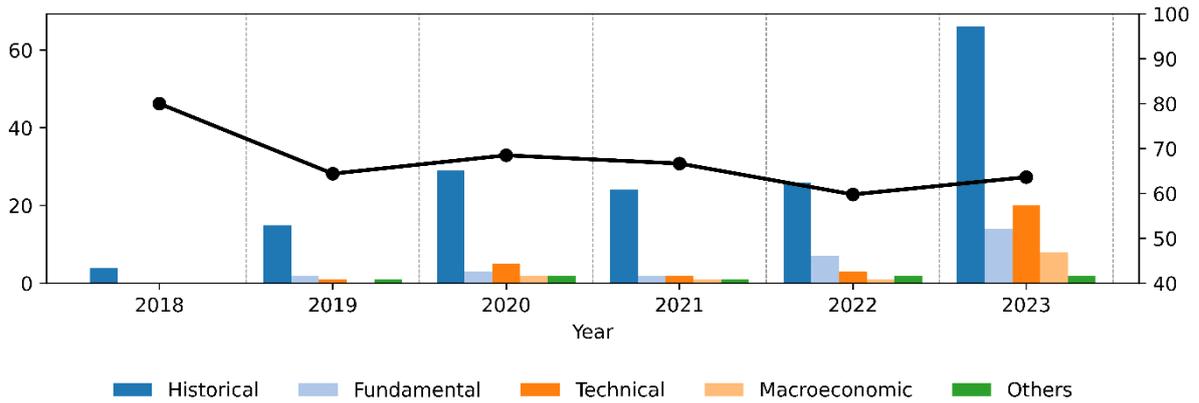

*Figure 18 - Data inputs used by year with journal filter (LHS). The line represents the percentage of articles using only Historical data (RHS).*

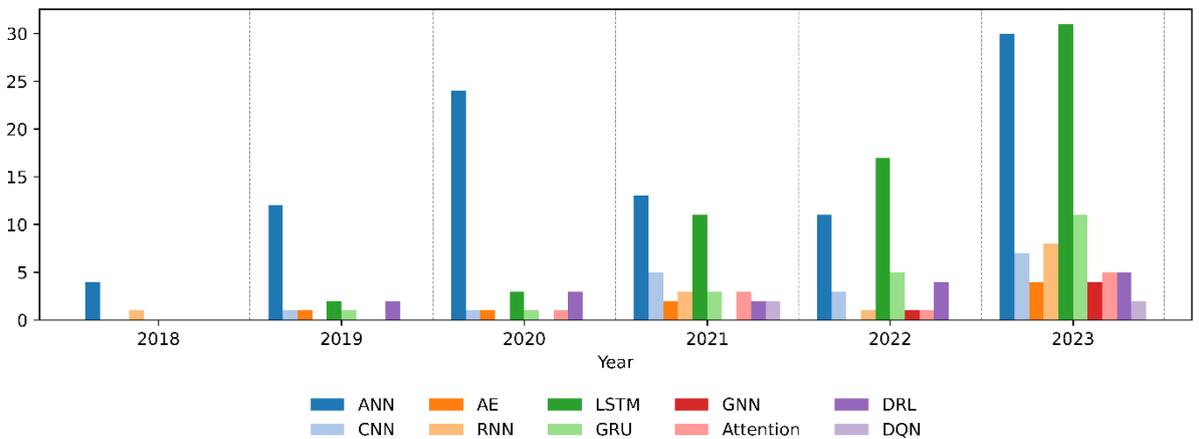

*Figure 19 - DL models with journal filter.*

The major difference in conclusion when applying the filter is shown using naïve models as a benchmark. The number of articles using naïve forecast methods in price forecast is 15.6% (up from 4.0%). In Portfolio management, the use of B&H (MW or EW) increased from 36.7% to 47.7%. In Algorithmic trading, articles incorporating Sharpe Ratio and MDD decrease from 62.5% to 50.0% and 39.6% to 35.7%, respectively. Moreover, 42.9% of the articles use B&H (MW or EW) as a benchmark, down from 45.8%.

Overall, we see a slight improvement in the best practices when applying a journal filter. However, these values remain small.

## 6 Conclusions

In conclusion, applying DL in financial asset management has showcased significant transformative potential. Enhanced predictive accuracy, integration of alternative data sources, and dynamic optimisation of portfolios are key benefits identified through multiple studies. Our extensive survey concluded that the number of articles within financial asset management has increased significantly over the past years, with a great focus on price forecasting. LSTMs, ANNs, and CNNs are still the preferred models by researchers, with attention mechanisms, GANs and GNNs gaining recent momentum.

From our study, we recommend four avenues for future research.

First, future research is needed to implement recent advances in computer science in all topics. In price forecast, the articles of Torres et al. (2021), Benidis et al. (2022), and X. Liu & Wang (2024) provide a good starting point. The authors survey recent advances in time series prediction, including in the scope of large language models, and these models still need to be implemented in financial asset management with possible better forecasts.

In value/factor investing, authors should focus on implementing explainable DL models. Models such as DeepLIFT (Shrikumar et al., 2017) and SHAP (Lundberg & Lee, 2017) have yet to be applied to this topic. DeepLIFT provides insight into the contribution of each input feature to a model's prediction by decomposing the prediction relative to a baseline, allowing investors to pinpoint critical factors. On the other hand, SHAP uses Shapley values from cooperative game theory to quantify the marginal impact of each feature on predictions consistently and fairly, offering a granular, feature-level explanation. Together, these methods enable investors to interpret black-box models in financial forecasting or stock scoring, ensuring alignment with fundamental principles of value investing. This increased level of explainability allows investors to understand the *why* behind the predictions, ultimately leading to more informed decisions.

Furthermore, in algorithmic trading and portfolio management, advances in DRL should also be considered (see X. Chen et al., 2023; Zhu et al., 2023). For example, Hierarchical DRL breaks down tasks, like asset selection and risk management, into smaller steps, making strategies more flexible. Multi-agent DRL models interactions between different market players or strategies, helping to simulate real market dynamics. Self-supervised DRL refines learning by finding patterns without needing detailed reward definitions, which is useful when defining success, which is tricky in finance.

Secondly, most studies focus mainly on a single type of asset, which limits how useful they are in different financial situations. It is important to include portfolios with multiple asset types and markets to make models more reliable and applicable worldwide. This supports the findings of Jiang (2021) and Cremers et al. (2019) and shows that research has remained the same over the last five years despite the growing complexity of global markets.

Thirdly, integrating alternative data inputs, such as sentiment analysis from social media and macroeconomic indicators, can significantly enhance the effectiveness of models, especially when dealing with multi-asset and multi-market portfolios. These data sources provide valuable insights into market sentiment and global economic trends, essential for understanding the interconnected behaviours of various asset classes and regional markets. For instance, macroeconomic indicators can help predict how broader economic changes impact diverse assets. By combining these data inputs, models can capture a more comprehensive picture of market dynamics, making them more adaptable and effective in managing complex, supporting multi-asset and multi-market portfolios.

Lastly, proper baseline and performance metrics should be standard in all studies for consistent comparisons and meaningful evaluations. While naïve baselines increased when journal criteria were applied in our analysis, these numbers remain limited. Establishing standardised baselines and metrics not only improves the reproducibility of studies but also facilitates meta-analysis. By providing source code through platforms like GitHub, authors can create opportunities for systematic meta-analysis, allowing researchers to compare methods, aggregate findings, and draw broader insights about the effectiveness and generalisability of models across diverse contexts. This collaborative effort can significantly advance the field.

In summary, with computer science's fast development, financial asset management still needs to catch up with recent developments. With this integration, we can provide evidence that potentially challenges the EMH and provide more robust evidence towards active versus passive portfolio management strategies.

Anadu, K., Kruttli, M., McCabe, P., & Osambela, E. (2020). The shift from active to passive investing: Risks to financial stability? *Financial Analysts Journal*, *76*(4), 23–39.

Barua, R., & Sharma, A. K. (2022). Dynamic Black Litterman portfolios with views derived via CNN-BiLSTM predictions. *Finance Research Letters*, *49*, 103111.

Behera, J., Pasayat, A. K., Behera, H., & Kumar, P. (2023). Prediction based mean-value-at-risk portfolio optimisation using machine learning regression algorithms for multi-national stock markets. *Engineering Applications of Artificial Intelligence*, *120*, 105843.

Belcastro, L., Carbone, D., Cosentino, C., Marozzo, F., & Trunfio, P. (2023). Enhancing Cryptocurrency Price Forecasting by Integrating Machine Learning with Social Media and Market Data. *Algorithms*, *16*(12), 542. https://doi.org/10.3390/a16120542

Benidis, K., Rangapuram, S. S., Flunkert, V., Wang, Y., Maddix, D., Turkmen, C., Gasthaus, J., Bohlke-Schneider, M., Salinas, D., Stella, L., & others. (2022). Deep learning for time series forecasting: Tutorial and literature survey. *ACM Computing Surveys*, *55*(6), 1–36.

Betancourt, C., & Chen, W.-H. (2021). Deep reinforcement learning for portfolio management of markets with a dynamic number of assets. *Expert Systems with Applications*, *164*, 114002.

Bustos, O., & Pomares-Quimbaya, A. (2020). Stock market movement forecast: A systematic review. *Expert Systems with Applications*, *156*, 113464.

Cao, J., Li, Z., & Li, J. (2019). Financial time series forecasting model based on CEEMDAN and LSTM. *Physica A: Statistical Mechanics and Its Applications*, *519*, 127–139. https://doi.org/10.1016/j.physa.2018.11.061

Cao, Y., Zha, D., Wang, Q., & Wen, L. (2023). Probabilistic carbon price prediction with quantile temporal convolutional network considering uncertain factors. *Journal of Environmental Management*, *342*, 118137.

Carhart, M. M. (1997). On persistence in mutual fund performance. *The Journal of Finance*, *52*(1), 57–82.

Ceffer, A., Fogarasi, N., & Levendovszky, J. (2018). Trading by estimating the quantised forward distribution. *Applied Economics*, *50*(59), 6397–6405.

Chan, J. Y.-L., Leow, S. M. H., Bea, K. T., Cheng, W. K., Phoong, S. W., Hong, Z.-W., Lin, J.-M., & Chen, Y.-L. (2022). A correlation-embedded attention module to mitigate multicollinearity: An algorithmic trading application. *Mathematics*, *10*(8), 1231.

Chen, B., Zhong, J., & Chen, Y. (2020). A hybrid approach for portfolio selection with higher-order moments: Empirical evidence from Shanghai Stock Exchange. *Expert Systems with Applications*, *145*, 113104.

Chen, X., Yao, L., McAuley, J., Zhou, G., & Wang, X. (2023). Deep reinforcement learning in recommender systems: A survey and new perspectives. *Knowledge-Based Systems*, *264*, 110335.

Cho, K. (2014). Learning phrase representations using RNN encoder-decoder for statistical machine translation. *ArXiv Preprint ArXiv:1406.1078*.

Cremers, K. J. M., Fulkerson, J. A., & Riley, T. B. (2019). Challenging the conventional wisdom on active management: A review of the past 20 years of academic literature on actively managed mutual funds. *Financial Analysts Journal*, *75*(4), 8–35.

Dakalbab, F., Talib, M. A., Nassir, Q., & Ishak, T. (2024). Artificial intelligence techniques in financial trading: A systematic literature review. *Journal of King Saud University-Computer and Information Sciences*, 102015.

DeMiguel, V., Garlappi, L., & Uppal, R. (2009). Optimal versus naive diversification: How inefficient is the 1/N portfolio strategy? *The Review of Financial Studies*, *22*(5), 1915–1953.

Devlin, J., Chang, M.-W., Lee, K., & Toutanova, K. (2018). *BERT: Pre-training of Deep Bidirectional Transformers for Language Understanding*.

Diebold, F. X., & Mariano, R. S. (2002). Comparing predictive accuracy. *Journal of Business & Economic Statistics*, *20*(1), 134–144.
29